\documentclass{mem}
\usepackage{natbib}\usepackage{txfonts}\usepackage{balance}
\usepackage{graphicx}
\usepackage[a4paper,breaklinks,dvipdfm]{hyperref}
\idline{75}{282}
\begin{document}
\def\teff{$T\rm_{eff }$}
\def\kms{$\mathrm {km s}^{-1}$}

\title{
Role of THESEUS in Understanding the Radiation Mechanism of GRB Prompt Emission
}

   \subtitle{}

\author{
R. \,Basak\inst{1,2} 
          }

\institute{
The Oskar Klein Centre for Cosmoparticle Physics, AlbaNova, SE-106 91 Stockholm, Sweden
\and
Department of Physics, KTH Royal Institute of Technology, AlbaNova University Center, SE-106 91 Stockholm, Sweden
\email{rupal.basak@gmail.com, rupalb@kth.se}
}

\authorrunning{Basak }

\titlerunning{GRB radiation study with THESEUS}

\abstract{
The radiation process of GRB prompt emission remains highly debated till date. 
Though a smoothly broken powerlaw function like Band provides an excellent fit 
to most of the cases, a spectrum with a broad top or double hump structure has 
emerged recently for several GRBs, specifically for the bright ones that have 
high signal to noise data. A number of models have been proposed to capture 
this shape which includes an additional blackbody component along with the 
Band, or a double smoothly broken powerlaw, or two blackbodies with a 
powerlaw/cut-off powerlaw and so on. However, finding the statistically 
favourable model has not always been possible primarily due to the limited resolution of 
GRB detectors. Recently, we have identified a number of interesting cases
where an observation with focusing detectors are available at a later part 
of the prompt emission. Their high resolution data helped in identifying additional
spectral component (a blackbody) in lower energies. 
The proposed \emph{THESEUS} mission, with its unprecedented sensitivity and 
spectral resolution available from the very beginning 
of the prompt emission holds the key to identify the correct spectral model and thereby 
the radiation process.

}
\maketitle{}

\section{Introduction}
Gamma-ray Bursts (GRBs) are the strongest sources of enormous gravitational 
and electromagnetic energy. From the recent detection of gravitational wave with 
electromagnetic counterpart \citep{gw_em},
we are now very sure that a class of GRBs known as the short GRBs are produced 
by coalescing binary compact stars in which at least one of the binary stars is 
a neutron star. On the other hand, the detection of core-collapse supernovae
during the afterglow of nearby long GRBs \citep{Galamaetal_1998, Hjorthetal_2003} points towards a collapsar scenario 
for the class of long GRBs. While the initial collapse or coalescence is the 
main source of gravitational wave, the electromagnetic energy is believed to 
be produced by a relativistic bipolar jet that is launched by the new born 
central engine. Yet, the mechanism by which the jet radiates remains a highly 
debated topic. Understanding the radiation process holds the key by which we 
can trace back the nature of the central engine which may not be determined 
by other methods. For example, in the discovery event of electromagnetic
counterpart involving coalescing neutron star, the nature of the end product
remains undetermined based on the mass estimate. The study of radiation process
may provide valuable input in this regard.

GRB radiation occurs in two phases -- prompt emission and afterglow. While 
the afterglow phase being a radiation from the shocked circumstellar matter
is related more to the external medium, the prompt emission is intrinsic 
to the GRB. Hence, in order to understand the GRB radiation process, it is 
necessary to model the profile of the prompt emission spectrum. However, the 
prompt emission spectroscopy faces a number of challenges as follows. 
(a) The spectrum evolves quite rapidly,
(b) There can be a number of overlapping pulses. Hence, it is difficult 
to follow the spectral evolution of the individual pulses \citep{Hakkila_Preece_2011}.
(c) GRB detectors are background dominated and their spectral sensitivity 
and resolution are quite poor. While the rapid spectral evolution demands 
fine time resolved study, the detectors are unable to provide good signal 
to noise ratio in the time resolved data.

In order to tackle these, a number of strategies have been adopted. For 
example, some studies involve only GRBs with single pulse \citep{Ryde_2004,
Ryde_Pe'er_2009}, and thus avoid 
the pulse overlapping effect. These studies generally find that the 
spectral evolution is quite smooth. Even for GRBs with multiple but 
separable pulses, the portion with no overlap do show a smooth evolution \citep{Basak_Rao_2013_parametrized}.
Some of these studies also show strong evidence of a thermal blackbody 
component mostly at the beginning of the prompt emission \citep{Ghirlandaetal_2003, Ryde_2004,
Rydeetal_2010_090902B}. A thermal 
component indicates an optically thick emission. In the standard fireball 
model, such an emission is in fact expected from the photosphere, 
though the actual spectral shape can be slightly broader compared to 
a blackbody due to sub-photospheric dissipations 
\citep{rees_and_meszaros_2005_prompt,Pe'er_2008, Pe'er_Ryde_2011}.

In recent years, we have adopted a new strategy to overcome the third 
problem \citep{Basak_Rao_2015_090618, Basak_Rao_2015_130925A}.
GRB detectors have wide field of view (FoV) in order to detect a GRB 
occurring at a random sky direction. But, this allows for huge background
and consequently the detectors are background limited and their sensitivity 
is quite poor. While it is necessary to have a wide FoV detector 
at the beginning, focusing instruments like \emph{Swift} X-ray Telescope (XRT)
can in principle operate at a later phase. These instruments have a much 
better sensitivity than the wide FoV detectors, and their CCD detectors have 
an order of magnitude better spectral resolution. Hence, they can be excellent
tools to find additional spectral component at lower energies with 
statistical significance. At the later phase, \emph{Swift} Burst Alert 
Telescope (BAT) type of detectors can also be employed to provide a wider 
energy coverage. But, \emph{Fermi} Gamma-ray Burst Monitor (GBM) cannot 
be used due to its limited sensitivity. The GBM can be used at the beginning 
in stead where the signal is higher, and GBM can provide a much wider 
energy coverage.

\begin{figure*}[t!]
\resizebox{\hsize}{!}{\includegraphics[clip=true]{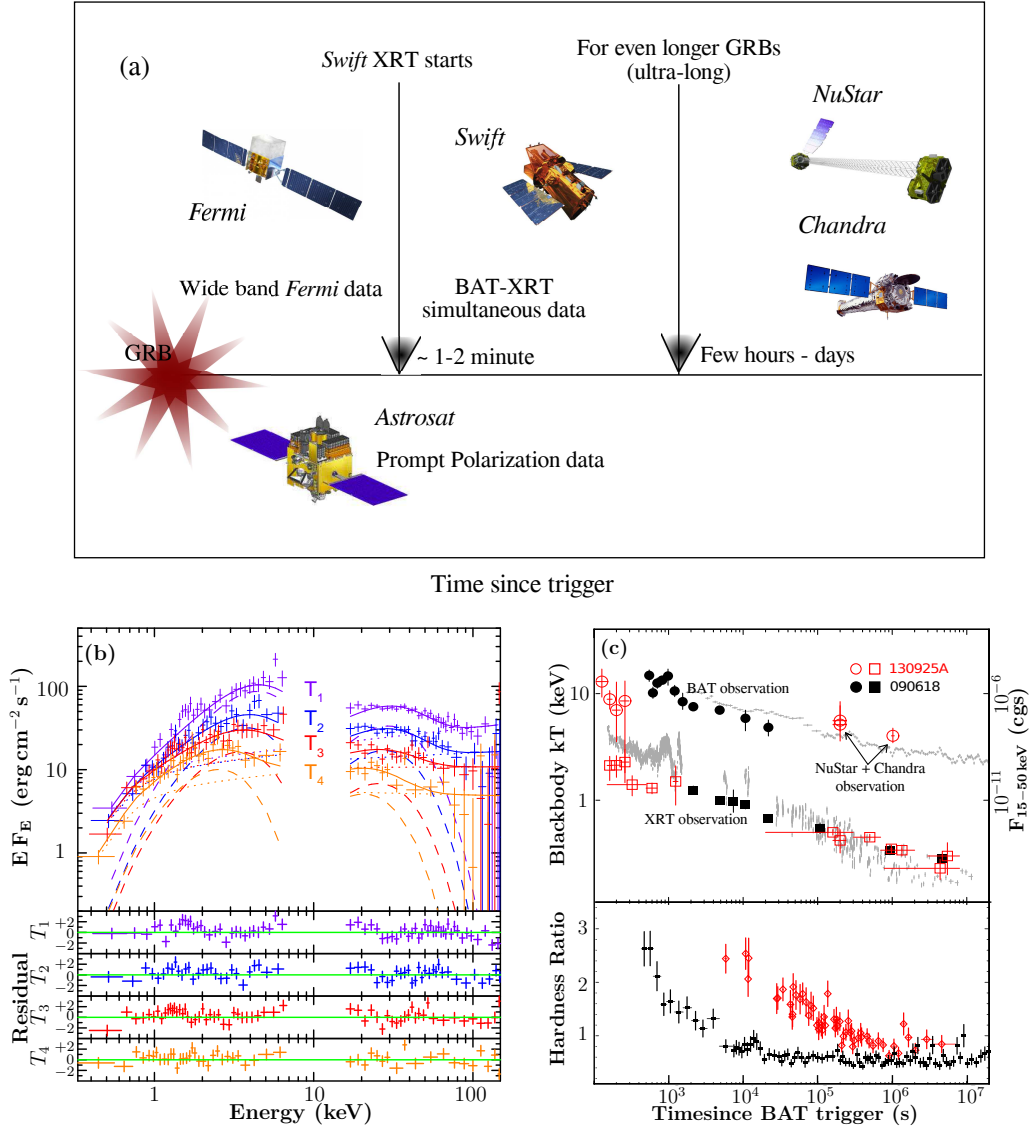}}
\caption{\footnotesize
\textit{(a):} The multi-instrument observation and analysis strategy is illustrated.
\textit{(b):} Two blackbodies and a powerlaw fit to the time resolved BAT-XRT data of GRB 
090618 \citep{Basak_Rao_2015_090618}. Upper panel shows the spectral fit with the two 
blackbodies shown by dashed line, and lower panel shows the fit residual (time ordered). 
\textit{(c):} Comparison of spectral evolution between GRB 130925A, an ultra-long GRB (open symbols) and 
GRB 090618, a long GRB (filled symbols). Upper panel shows the evolution of the 
temperature. The lightcurve is shown in the background (higher one for 090618).
Lower panel shows the hardness ratio which is defined as the count ratio of
1.5--10\,keV band to 0.3-1.5\,keV band.
\citep{Basak_Rao_2015_130925A}.
}
\label{multi_instrument}
\end{figure*}

In addition to this, the CZT imager (CZTI) of the Indian space mission 
\emph{Astrosat} is providing the valuable polarization measurement during 
the prompt emission. A joint spectro-polarimetric study is first of its 
kind. As different model of prompt emission predict different 
degree of polarization, this measurement can give independent constraints  
on the possible model. The strategy outlined above is illustrated in 
Fig.~\ref{multi_instrument}\,a.

The proposed \emph{THESEUS} mission \citep{theseus2017} 
will have an enormous capability 
to automate the strategy of obtaining high resolution data in a wide band,
and that too from the very beginning of the prompt emission. Hence, the 
above strategic observation can be routinely done without human intervention.
In this paper, we highlight some of the strategic analysis and how 
\emph{THESEUS} can contribute in studying GRB prompt emission spectrum.

\begin{table*}
\caption{Sample of the strategic analysis}
\label{sample}
\begin{center}
\begin{tabular}{ccccccc}
\hline
\\
GRB & \emph{Fermi} GBM & \emph{Fermi} LAT & \emph{Swift} BAT & \emph{Swift} XRT & Other & Polarization\\
\hline
\\
090618 & Yes & No & Yes & Yes & & No \\
130925A & No & No & Yes & Yes & \emph{Nustar}, \emph{Chandra} & No \\
151006A & Yes & Yes & Yes & Yes & & Yes \\
\\
\hline
\end{tabular}
\end{center}
\end{table*}

\section{Strategic Analysis}
We demonstrate the strategic analysis with three GRBs namely, GRB 
090618, GRB 130925A and GRB 151006A. 

GRB 090618 is a bright GRB detected by both \emph{Fermi} and \emph{Swift}. 
The prompt emission lasted quite long ($\sim200$\,s), and the XRT started
observing the GRB from 125\,s after the BAT trigger time. The XRT covers the 
falling part of the last pulse of the prompt emission. As this part is 
devoid of pulse overlapping effect, the spectrum is expected to show a 
smooth evolution. \cite{Pageetal_2011}, using the XRT data at this phase
find an evidence of a thermal component. The temperature and the normalization 
of this component decreases monotonically and smoothly with time. 
In our analysis of this GRB \citep{Basak_Rao_2015_090618}, see Fig.~\ref{multi_instrument}\,b, 
we include the BAT detector at the late phase and further divide the 
BAT-XRT overlapping time into four equal time bins. The joint BAT-XRT data 
shows that there is an evidence of a blackbody component in the XRT band 
as pointed out by \cite{Pageetal_2011}, while in the BAT energy band another
blackbody component shows up, see Fig.~\ref{multi_instrument}\,b. The 
temperature of these two blackbodies evolve in the same manner. For the data 
before and after the joint BAT-XRT observation, we fit a single blackbody.
The temperature of the single blackbody fitted to the BAT only data in the 
previous bins show a similar evolution as that of the higher-temperature
blackbody of the joint BAT-XRT data. Also, the temperature of a single 
blackbody fitted to the XRT data at later bins show a similar evolution 
as the lower-temperature blackbody of the joint data. This indicates that 
the spectrum always has the double hump modelled with two blackbodies, but 
they show up only when we have a good enough spectral coverage with BAT and 
XRT. This example illustrates the importance of the high resolution XRT data 
at the right moment of GRB emission. 

GRB 130925A is an interesting and a highly debated event. First of all, was it 
at all a GRB or a tidal disruption event (TDE)? Though an image taken with Hubble Space Telescope 
\citep{tanvir} shows an off-set from the host galaxy, but as the morphology 
of the host points to a recent major merger, it cannot fully rule out a TDE.
Hence, it is very important to study the radiation process and compare with other 
GRBs. If the event is a GRB, it is the longest ultra-long GRB till date. 
The event got considerable attention and was observed as a target of opportunity 
(ToO) with \emph{Nustar} and \emph{Chandra} at a later phase, see Table~\ref{sample}.
The radiation process of the event was also debated. \cite{Piroetal_2014} using 
the XRT data till late time find an evolving blackbody with a temperature 
varying from $\sim1$\,keV to $\sim0.5$\,keV. But, \cite{nustar_2014} using 
the \emph{Nustar} and \emph{Chandra} data find a temperature at a much 
higher value $\sim5$\,keV. These two observations already indicates the presence
of two backbodies. When we combine the data and also include the BAT data at the 
beginning, we find two blackbody components throughout the burst with the temperature
falling off quite smoothly with time \citep{Basak_Rao_2015_130925A}. Moreover, when 
we compare the evolution of the temperature and hardness ratio of this event 
with that of the long GRB 090618, we find a remarkable similarity, see 
Fig.~\ref{multi_instrument}\,c. The only difference is the time scale of variation 
which is much slower in the ultra-long case denoting a much larger progenitor 
that underwent a collapse. Hence, our observation indicates that first of all, 
this event is a GRB and second, the radiation process of the ultra-long GRB is 
also similar to the long GRB class.

Our third example is GRB 151006A \citep{Raoetal_2016, Basaketal_2017}. This is 
the first GRB detected by \emph{Astrosat} CZTI and has a polarization measurement
\citep{Raoetal_2016}. We fit a Band function to the time resolved spectra. We 
find that the peak of the spectrum initially shows a hard-to-soft evolution, 
but then show a sudden jump reaching values beyond MeV. The changeover time 
also coincide with the first LAT detected photon. This indicates either a beginning
of an afterglow or a second pulse of the prompt emission, which however is not 
seen in the lightcurve at any energy. Our analysis show that it is indeed the 
second case because (a) the late time data from BAT and XRT shows considerable 
curvature in the data unlike the simple powerlaw shape of afterglow spectrum,
(b) the hardness ratio plot shows an underlying second pulse at this phase,
(c) also the polarization data at these two phases are quite high which is unattainable 
by the random magnetic field expected during an afterglow phase. This 
study shows the power of the strategic analysis along with polarimetric data.
The CZTI has by now detected over 50 GRBs and in several cases we have polarization
measurement \citep{Chattopadhyayetal_2017}. 
Over the mission lifetime we expect to observe at least 50 GRBs with measured 
polarization, which will be a huge contribution in understanding the correct radiation 
process.

%

\section{Role of THESEUS}
The proposed \emph{THESEUS} mission \citep{theseus2017} has a number of unique capabilities that will 
be helpful in the study of prompt emission spectrum. 

The three units of X-Gamma Imaging Spectrometer (XGIS) cover a wide energy range of 
2 keV -- 20 MeV. The detector is a combination of solid state detector -- a 
Si drift detector, and a scintillator -- CsI (Tl). The detector provides 
an unprecedented resolution of 200 eV FWHM at 6 keV in the energy range 2--30\,keV.
The resolution at higher energies are moderate -- 18\% FWHM at 60 keV (30--150\,keV)
and 6\% FWHM at 500 keV ($>150$\,keV). This instrument has a coded aperture mask (CAM)
as used in case of \emph{Swift} BAT. It has a large enough FoV ($64^{\circ}\times64^{\circ}$ 
full FoV in the range 2--30\,keV and $2\pi$\,sr at energies $>150$\,keV) to 
detect GRBs. 

In addition, the Soft X-ray Imager (SXI) of \emph{THESEUS} with its four ``Lobster-Eye''
provides a FoV of 1\,sr and can detect the prompt emission in the softer energies. 
The CCD covers an energy range 0.3--6\,keV which will complement the range covered 
by XGIS. This energy range is also important to constraint the line of sight column 
density that must be included in the analysis of soft X-ray data. 

The high resolution detectors that are capable of detecting GRBs makes the \emph{THESEUS}
an extremely powerful mission for prompt emission spectroscopy. As this high resolution 
data at lower energies is available from the very beginning of the prompt emission,
the strategic analysis described above will be routinely done by \emph{THESEUS}.

\section{Conclusions}
The prompt emission of GRBs remains a highly debated topic. The main hurdle seems 
to be the quality of data that can be afforded maintaining a large FoV of the 
GRB instruments. The detectors in general suffer from poor sensitivity and 
resolution. We have shown that meaningful spectral evolution studies can be 
performed if we take certain strategy in our analysis. These includes e.g.,
high resolution data at the late phase of the prompt emission corroborated 
with the wide band data from \emph{Fermi} at the beginning of the GRB. 
The polarization measurement with \emph{Astrosat} CZTI adds another dimension
to this effort. 

The proposed mission \emph{THESEUS} has a wide enough FoV, and the GRB 
detectors have an order of magnitude better spectral resolution at lower energies 
compared to the traditional GRB detectors. Moreover, this high resolution data 
is available from the very beginning of the prompt emission. Hence, it is 
expected that the prompt emission spectroscopy will get a boost with the 
advent of this type of mission.

\begin{acknowledgements}
We thank A. R. Rao and the CZTI collaboration for the contribution in the 
published works. We gratefully acknowledge L. Amati and other members of 
the \emph{THESEUS} collaboration for the information regarding the mission.
\end{acknowledgements}

\bibliographystyle{aa}

\end{document}